\documentclass[aps,prl,twocolumn]{revtex4-1}
\usepackage{amsmath}
\usepackage{amscd}
\usepackage{wasysym}
\usepackage[T1]{fontenc}
\usepackage{ae,aecompl}
\usepackage{amsfonts}
\usepackage{amsthm}
\usepackage{dsfont}
\usepackage{graphicx}
\usepackage{amsfonts}
\graphicspath{{./figs/}}
\newcommand{\ue}{\mathrm{e}}

\newcommand{\la}{\langle}
\newcommand{\ra}{\rangle}

\newcommand{\R}{\mathbb{R}}

\begin{document}
\title{Non-adiabatic transitions through exceptional points in the band structure of a PT-symmetric lattice}
\author{Bradley Longstaff and Eva-Maria Graefe}
\address{Department of Mathematics, Imperial College London, London, SW7 2AZ, United Kingdom}
\begin{abstract}
Exceptional points, at which two or more eigenfunctions of a Hamiltonian coalesce, occur in non-Hermitian systems and lead to surprising physical effects. In particular, the behaviour of a system under parameter variation can differ significantly from the familiar Hermitian case in the presence of exceptional points. Here we analytically derive the probability of a non-adiabatic transition in a two-level system driven through two consecutive exceptional points at finite speed. The system is Hermitian far away from the exceptional points. In the adiabatic limit an equal redistribution between the states coalescing in the exceptional point is observed, which can be interpreted as a loss of information when passing through the exceptional point. For finite parameter variation this gets modified.  We demonstrate how the transition through the exceptional points can be experimentally addressed in a PT-symmetric lattice using Bloch oscillations. 
\end{abstract}
\maketitle

\section{Introduction}
The intriguing properties of open quantum systems described by non-Hermitian and PT-symmetric Hamiltonians have opened up a new area of research \cite{Bend18_book,Chris18_book}. The difference between Hermitian and non-Hermitian quantum physics is most pronounced in the presence of exceptional points, at which two or more eigenstates of a system coalesce. They lead to a number of counterintuitive features and have attracted a large amount of research interest, both theoretically and experimentally \cite{Heis12,Brod13,Menk16,Thom16,Chen17,Hoda17,Lupu17,Achi17,Gold18,Long19}. A number of these studies have investigated the behaviour of the wave function when parameters are varied cyclically around such a point \cite{Dopp16,Xu16,Hass17,Zhan18}. In Hermitian quantum physics the adiabatic theorem ensures that a system prepared in an eigenstate remains in an instantaneous eigenstate, when parameters are varied sufficiently slowly. The situation is more involved in non-Hermitian systems. Here the exponential relative decay between different eigenstates competes with the exponentially small non-adiabatic corrections, which can lead to apparently non-adiabatic behaviour even in the adiabatic limit of infinitely slow parameter variation \cite{Berr11,Uzdi11,Grae13b,Milb15}. This causes asymmetric behaviour when encircling exceptional points. 

Here we go one step further and consider the behaviour of a system driven directly through an exceptional point. In particular, we study a two-level system driven through two consecutive exceptional points, at finite speed and in the adiabatic limit. The system is close to being Hermitian at the beginning and the end of the parameter sweep, making it meaningful to study the ratio of transmitted population between the instantaneous eigenstates far away from the exceptional points. We derive an {\it analytic} expression for the transmission probability. For adiabatic parameter variations this predicts a loss of information, leading to an equal redistribution of the population between the states coalescing in the exceptional point. In the fast driving limit quantum quench behaviour is recovered.  We demonstrate how this could be observed using Bloch oscillations in a PT-symmetric lattice in a realistic experimental setup. Interestingly, the Hamiltonian of this system itself does not have any exceptional points. A similar effect of partial transitions between Bloch bands at exceptional points has recently been observed numerically in a more complicated lattice structure and experimentally using optical fibre loops \cite{Bend15,Wimm15}. The quantitative description provided here explains these effects and opens up new avenues for the control of optical beams. 

\section{Non-adiabatic transitions in a two-mode system}
Consider the PT-symmetric Hamiltonian
\begin{equation}
\label{eqn:LZHam}
\hat H= \begin{pmatrix} -v & i\gamma \\ i\gamma & v \end{pmatrix},
\end{equation}
where $\gamma$ and $v$ are real parameters and we set $\gamma>0$ without loss of generality. This Hamiltonian describes two states with an energy difference of $2v$ and an asymmetric coupling $i\gamma$. Its direct implementation in a two-waveguide setup is nontrivial, due to the nonreciprocal coupling between the two modes. Nonreciprocal coupling between two resonators has been discussed for example in \cite{Long15b}. In this work we shall consider a perhaps slightly less obvious implementation of the Hamiltonian (\ref{eqn:LZHam}), as the Bloch Hamiltonian of a PT-symmetric tight-binding lattice. This setup allows for a direct visual observation of the population transfer between the two modes when the system is driven through the exceptional points. 

\begin{figure}
\begin{center}
\includegraphics[width=0.238\textwidth]{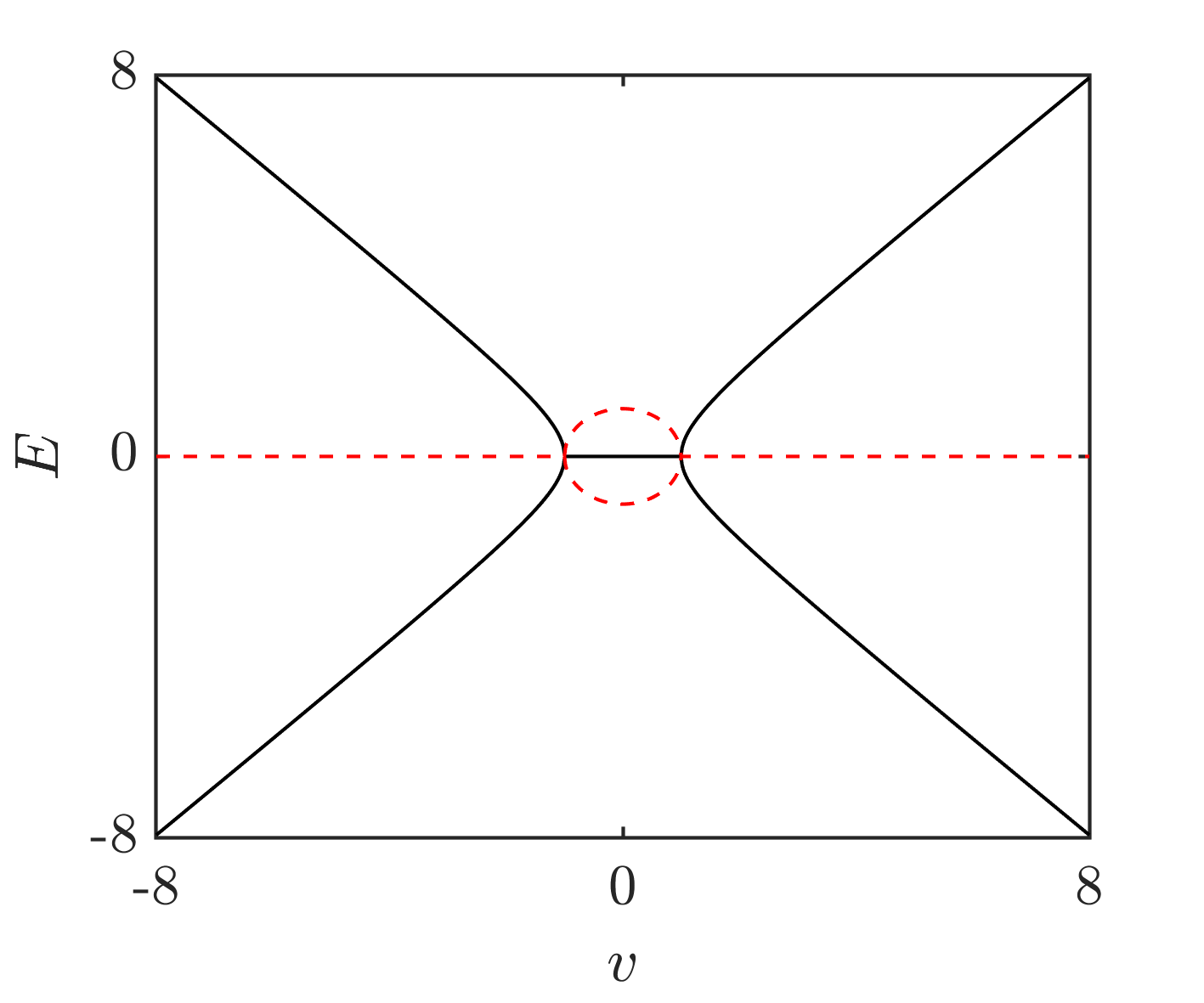}
\includegraphics[width=0.238\textwidth]{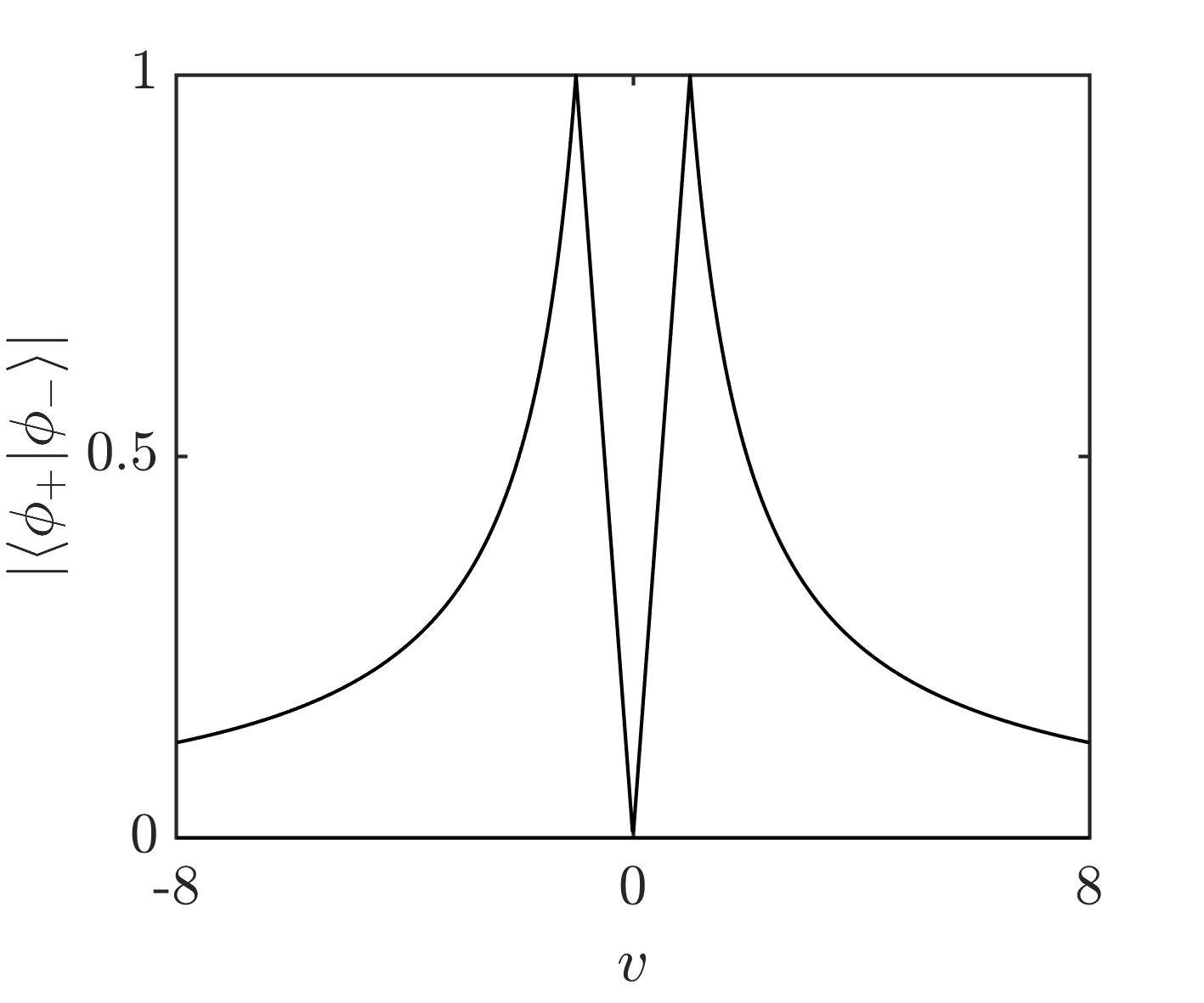}
\end{center}
\caption{\label{fig_eigval1} Eigenvalues and overlap of the eigenstates of the model Hamiltonian (\ref{eqn:LZHam}). The left panel shows the real (solid black line) and imaginary (dashed red line) parts of the eigenvalues, the right panel shows the overlap of the eigenstates, both in dependence on the coupling strength $v$ for a fixed value of $\gamma=1$.}
\end{figure}

The eigenvalues of the Hamiltonian (\ref{eqn:LZHam})
\begin{equation}
\lambda_{\pm}=\pm\sqrt{v^2-\gamma^2}
\end{equation}
in dependence on $v$ for a fixed value of $\gamma$ are depicted in the left panel of Fig. \ref{fig_eigval1}.  The right panel shows the overlap between the two (right) eigenstates in dependence on $v$. For values of $|v|>\gamma$ the eigenvalues are real and for very large values of $|v|$ the eigenstates are almost orthogonal (and tend towards the standard basis vectors), leading the system to behave in an essentially Hermitian way. For intermediate values of $v$ the non-Hermiticity is more apparent in the non-vanishing overlap of the eigenstates. The system has two exceptional points located at $v=\pm \gamma$, at which both eigenvalues coalesce and the Hamiltonian has only a single eigenstate. For $|v|<|\gamma|$ the eigenvalues are complex conjugate and there is one exponentially growing and one exponentially decaying mode. 

Suppose that the system is initially in one of the eigenstates, for very large negative $v\ll-\gamma$, and $v$ is increased adiabatically. One expects the state to closely follow the instantaneous eigenstate it originated from up until the exceptional point. At this point the initial instantaneous eigenstate coalesces with the other and it cannot be inferred in which of the two states the system originated. Thus, the population of the two instantaneous eigenstates is expected to be equal immediately after the exceptional point. However, this argument does not take into account that once the parameters have passed through the first exceptional point, one of the states decays exponentially while the other grows exponentially. If $v$ is varied adiabatically, the system has enough time to ``switch" to the configuration where the entire remaining population is in the exponentially growing state. When the system later passes through the second exceptional point the population is again equally distributed between the two eigenstates. Now both the states are stable and for large positive values of $v$ the population is expected to be equally distributed between the two instantaneous eigenstates. That is, information about the initial state is lost when the system is driven {\it adiabatically} through the exceptional points. After driving the system though the pair of exceptional points the state of the system is given by $\tilde{\psi} = (\tilde{\psi}_1,\tilde{\psi}_2)$ in the basis of eigenstates. In the adiabatic limit we have $|\tilde{\psi}_1|^2 = \frac{1}{2} = |\tilde{\psi}_2|^2$. Thus, while the initial state has two degrees of freedom (for example the relative phase between the two components and an amplitude), the final state is parameterised by a single variable, which is a relative phase between the two components. As there is clearly no one-to-one mapping from a one-dimensional space to a two-dimensional space, one cannot time-reverse $\tilde{\psi}$ to obtain the initial state. Of course, this observation is rather academic in nature, as any experiment will not be truly adiabatic. Thus, one might in principle be able to obtain the initial state by time reversing the dynamics. However, our numerical simulations show that the backwards time evolution can be very sensitive to small perturbations, which could make it practically very difficult to time-reverse the final state to obtain the initial state.

\begin{figure}
\begin{center}
\includegraphics[width=0.238\textwidth]{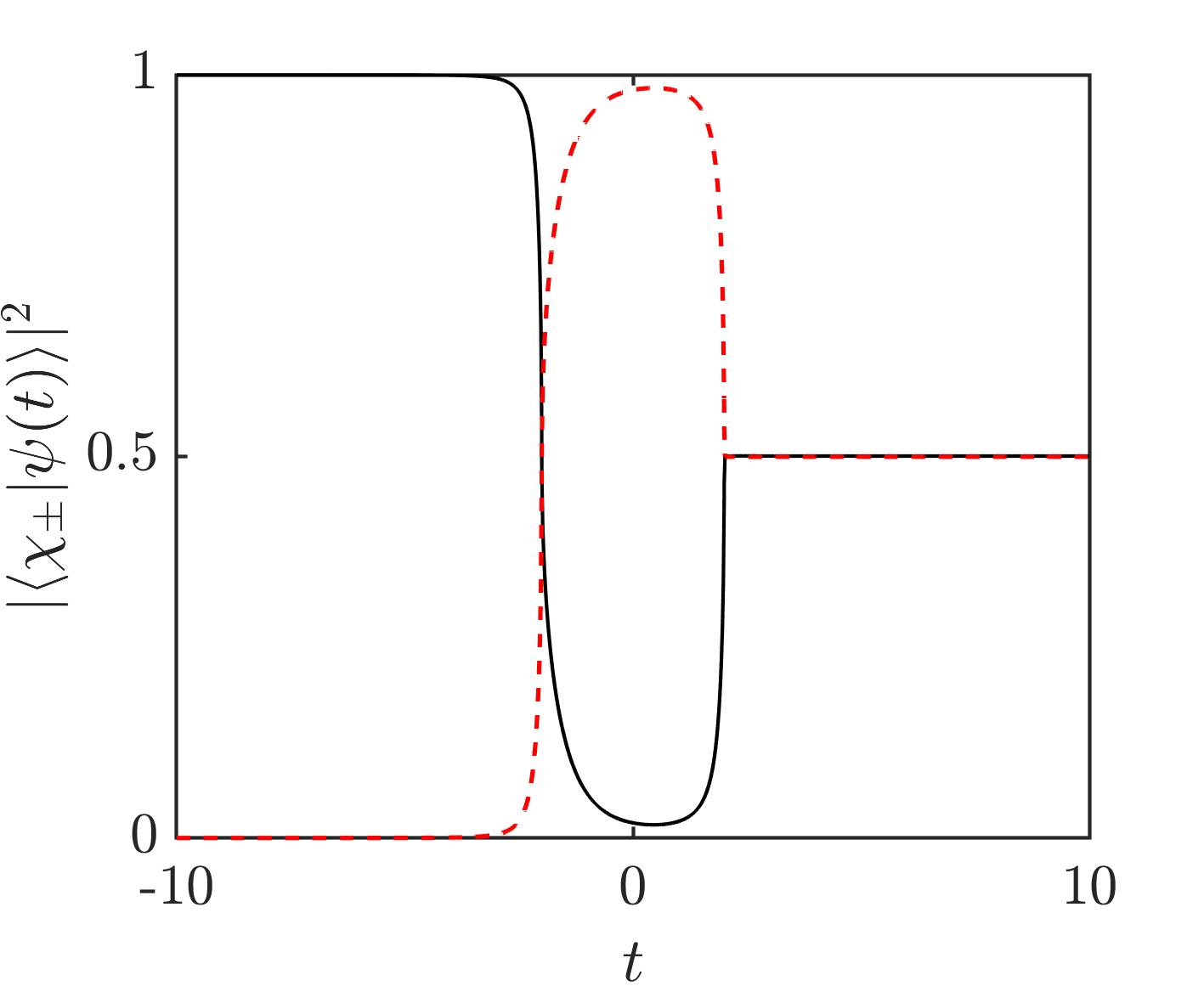}
\includegraphics[width=0.238\textwidth]{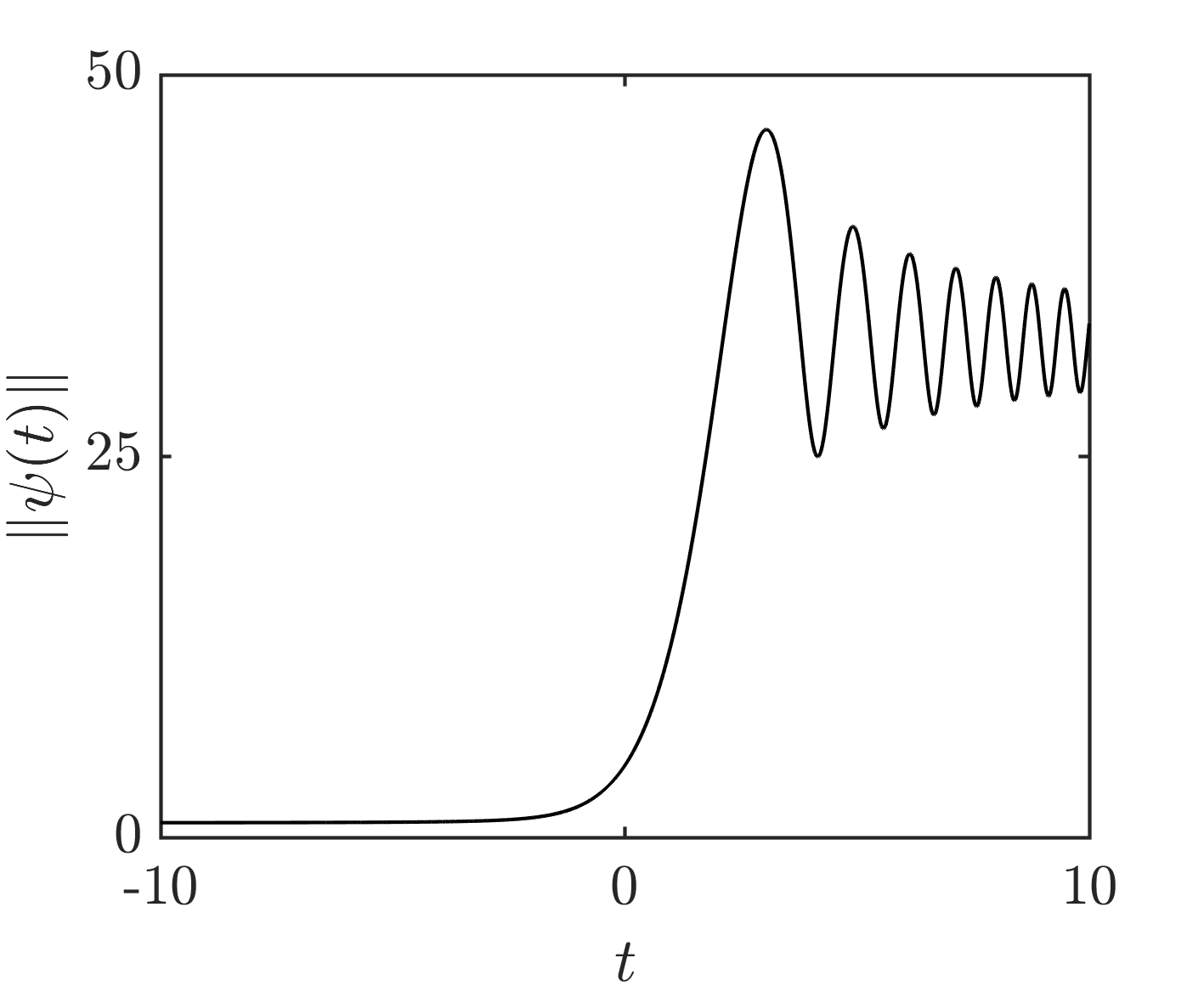}
\includegraphics[width=0.238\textwidth]{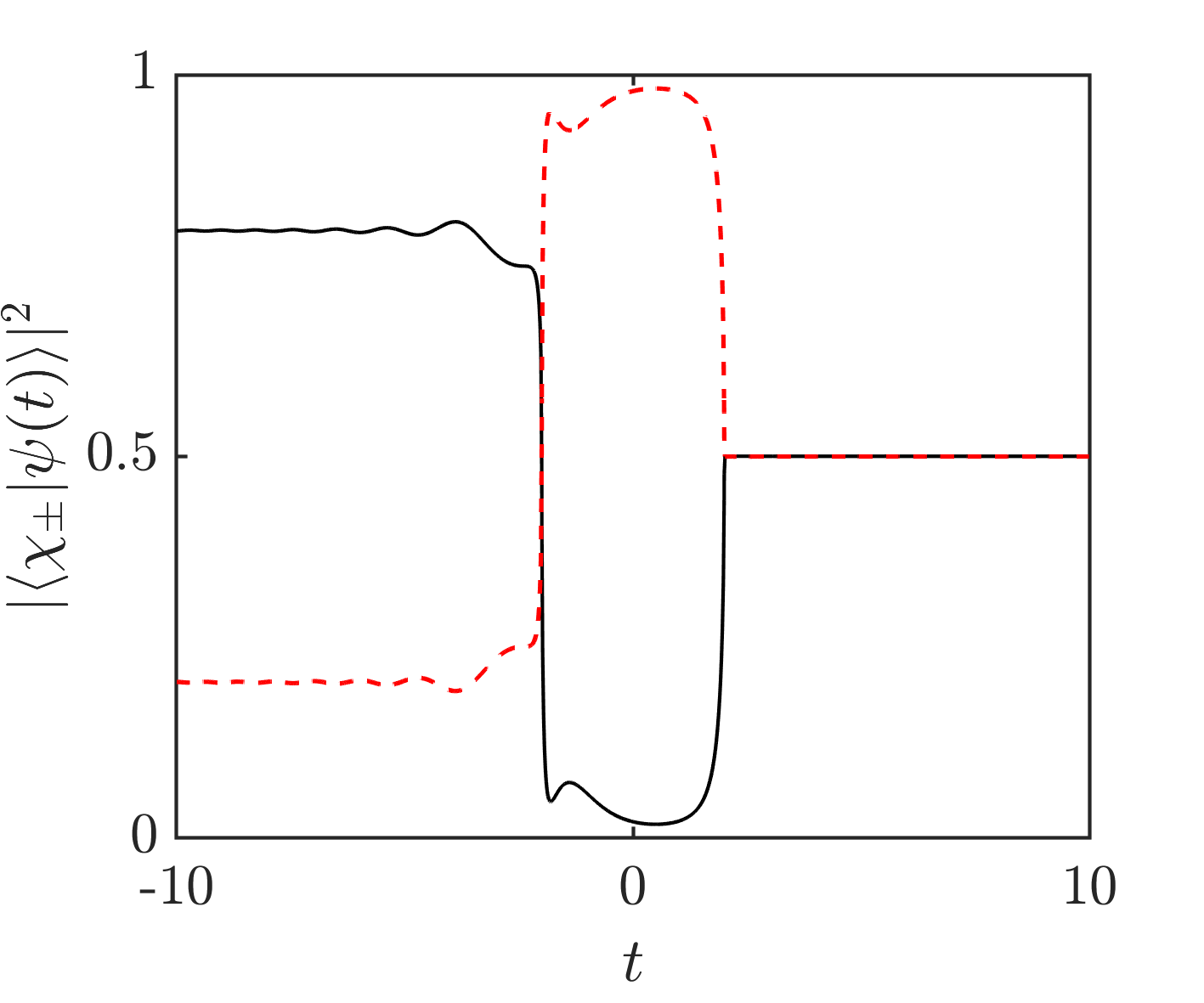}
\includegraphics[width=0.238\textwidth]{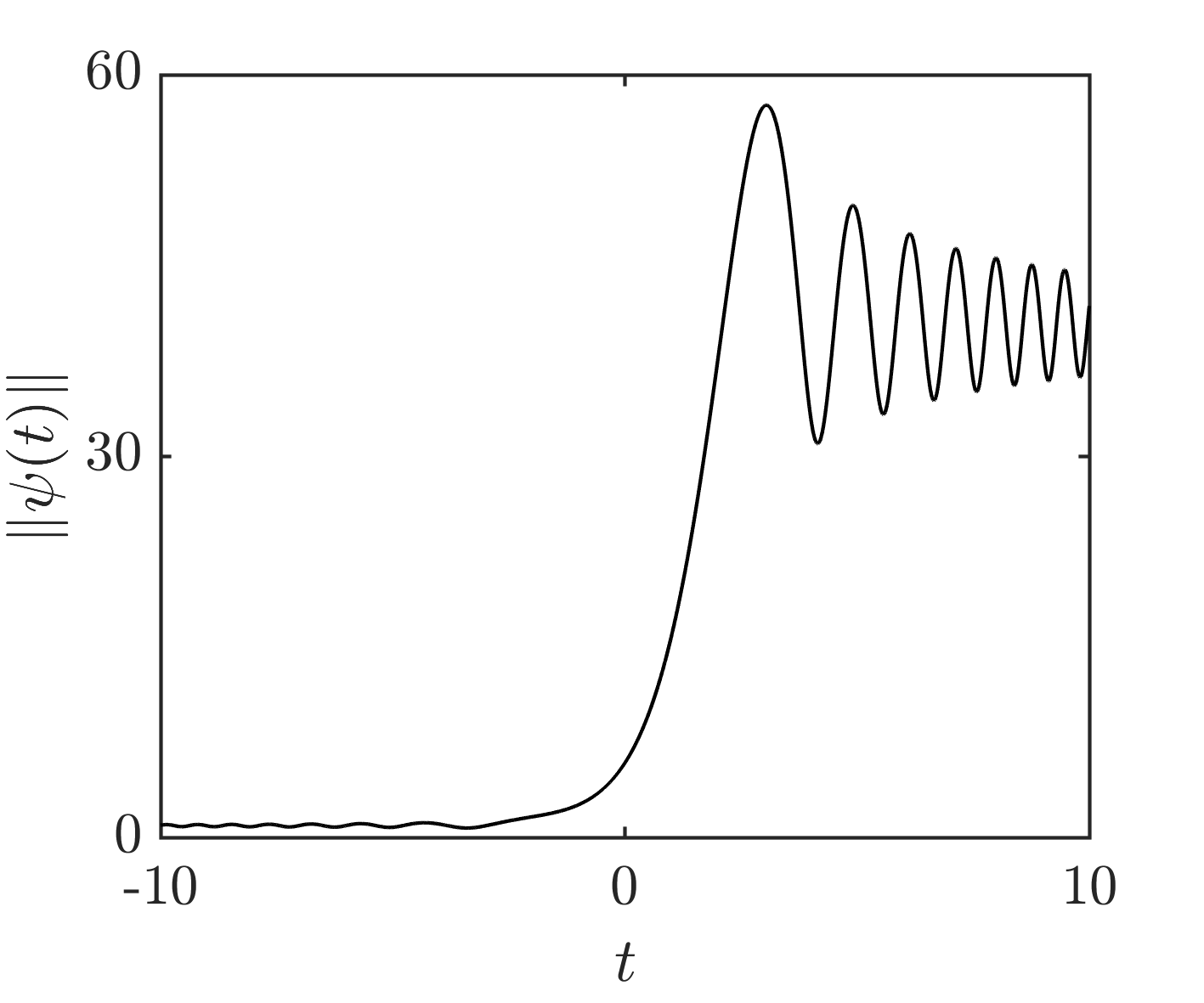}
\end{center}
\caption{\label{fig_adiab_dyn1} Dynamics generated by the Hamiltonian (\ref{eqn:LZHam}). $v$ is varied linearly in time as $v=\alpha t$ from $v_{i}=-5$ to $v_{f}=5$ with $\alpha=0.5$ and $\gamma=1$. The initial state is one of the instantaneous eigenstates (top) and a randomly selected state (bottom). The left panel shows the population in the two instantaneous eigenstates, the right panel shows the overall norm of the wave function.}
\end{figure}

This behaviour is indeed observed in numerical simulations and occurs independently of the initial state. In Fig. \ref{fig_adiab_dyn1} the relative population of the instantaneous eigenstates (calculated as the normalised projection onto the left eigenstate) and the overall norm of the wave function are plotted as a function of time. Here $v=\alpha t$ is slowly varied from a large negative initial value to a large positive final value for an initial eigenstate (top) and a randomly selected initial state (bottom). The norm, depicted on the right,  grows approximately exponentially in the region between the two exceptional points and continues to oscillate after the second exceptional point due to the lack of orthogonality of the eigenstates. The oscillations die off at large values of $v$ where the eigenstates become approximately orthogonal. 

In practice, however, parameters are not always varied slowly, and we are thus interested in the behaviour of the system when $v$ is varied non-adiabatically. This can be understood analytically by closely following Zener's derivation of the famous Landau-Zener-Majorana-St\"uckelberg formula \cite{Zene32}. We assume the parameter $v$ to vary linearly in time, that is, $v=\alpha t$, $\alpha \in \R^+$, where $t$ runs from minus to plus infinity.  At $t\to\pm\infty$ the eigenstates are given by the two uncoupled levels, i.e. the standard basis. We assume that the system is initially in the eigenstate 
\begin{equation}
\label{eqn:boundary conds}
|\psi_1\left(t \to -\infty\right)|^2 = 0, \quad |\psi_2\left(t \to -\infty\right)|^2 = 1.
\end{equation} 
We want to deduce the transmission probability into the same diabatic state at $t\to+\infty$, given by
\begin{equation}
\label{eqn:landau zener def}
P_{\rm tr} = \frac{|\psi_2\left(t \to +\infty\right)|^2}{|\psi_1\left(t \to +\infty\right)|^2+|\psi_2\left(t \to +\infty\right)|^2}.
\end{equation}

For this purpose we start from the Schr\"odinger equation for $\psi = \left(\psi_1,\psi_2\right)$
\begin{align}
i \dot{\psi}_1 &= -\alpha t \psi_1 + i\gamma \psi_2, \label{eqn:psi1 eom}\\
i\dot{\psi}_2 &= i\gamma \psi_1 + \alpha t \psi_2, \label{eqn:psi2 eom}
\end{align}
and transform (\ref{eqn:psi1 eom}) into the second order differential equation
\begin{equation}
\label{eqn:psi1 second}
\ddot{\psi}_1 + \left(-\gamma^2 - i\alpha + \left(\alpha t\right)^2\right)\psi_1 = 0.
\end{equation}
Applying the transformation $z(t) = \ue^{-i\pi/4}\sqrt{2\alpha} t$ converts this equation into the Weber equation 
\begin{equation}
\label{eqn:psi1 weber}
\frac{d^2 \psi_1}{dz^2} + \left(\nu + \frac{1}{2} - \frac{1}{4}z^2\right)\psi_1 = 0,
\end{equation}
with $\nu = -i\beta$ and $\beta = \gamma^2/2\alpha$.
A solution satisfying the initial conditions (\ref{eqn:boundary conds}) is given by
\begin{equation}
\label{eqn:phi1 soln}
\psi_1(t) = A D_{-\nu - 1} \left(-iz(t)\right),
\end{equation}
where $A$ is a normalisation factor and $D_\nu$ is a Weber (or parabolic cylinder) function \cite{Abra72}. The normalisation factor is determined from the asymptotic value
\begin{equation}
\label{eqn:phi1 asymp m}
\psi_1\left(t \to -\infty\right) = A \ue^{-i\pi \left(\nu+1\right)/4}\ue^{-iR^2/4}R^{-\nu - 1},
\end{equation}
with $R = \sqrt{2\alpha}t$. Inserting this into the equation of motion (\ref{eqn:psi1 eom}) provides an asymptotic expression for $\psi_2$
\begin{equation}
\label{eqn:phi2 asymp m}
\psi_2\left(t \to -\infty\right) = -i\frac{\sqrt{2\alpha}}{\gamma}A\ue^{-i\pi \left(\nu + 1\right)/4}\ue^{-iR^2/4}R^{-\nu},
\end{equation}
which, together with the initial conditions, yields $|A|^2 = \beta \ue^{\pi \beta/2}$. Making use of the asymptotic value
\begin{equation}
\psi_1\left(t \to +\infty\right) = A \frac{\sqrt{2\pi}}{\Gamma\left(\nu + 1\right)}\ue^{i\pi\nu/4}\ue^{iR^2/4}R^\nu,
\end{equation} 
and well-known properties of the gamma function, leads to the amplitude
\begin{equation}
\label{eqn:phi1 mod asymp}
|\psi_1\left(t \to +\infty\right)|^2 = \ue^{2\pi\beta}-1.
\end{equation}

\begin{figure}
\begin{center}
\includegraphics[width=0.238\textwidth]{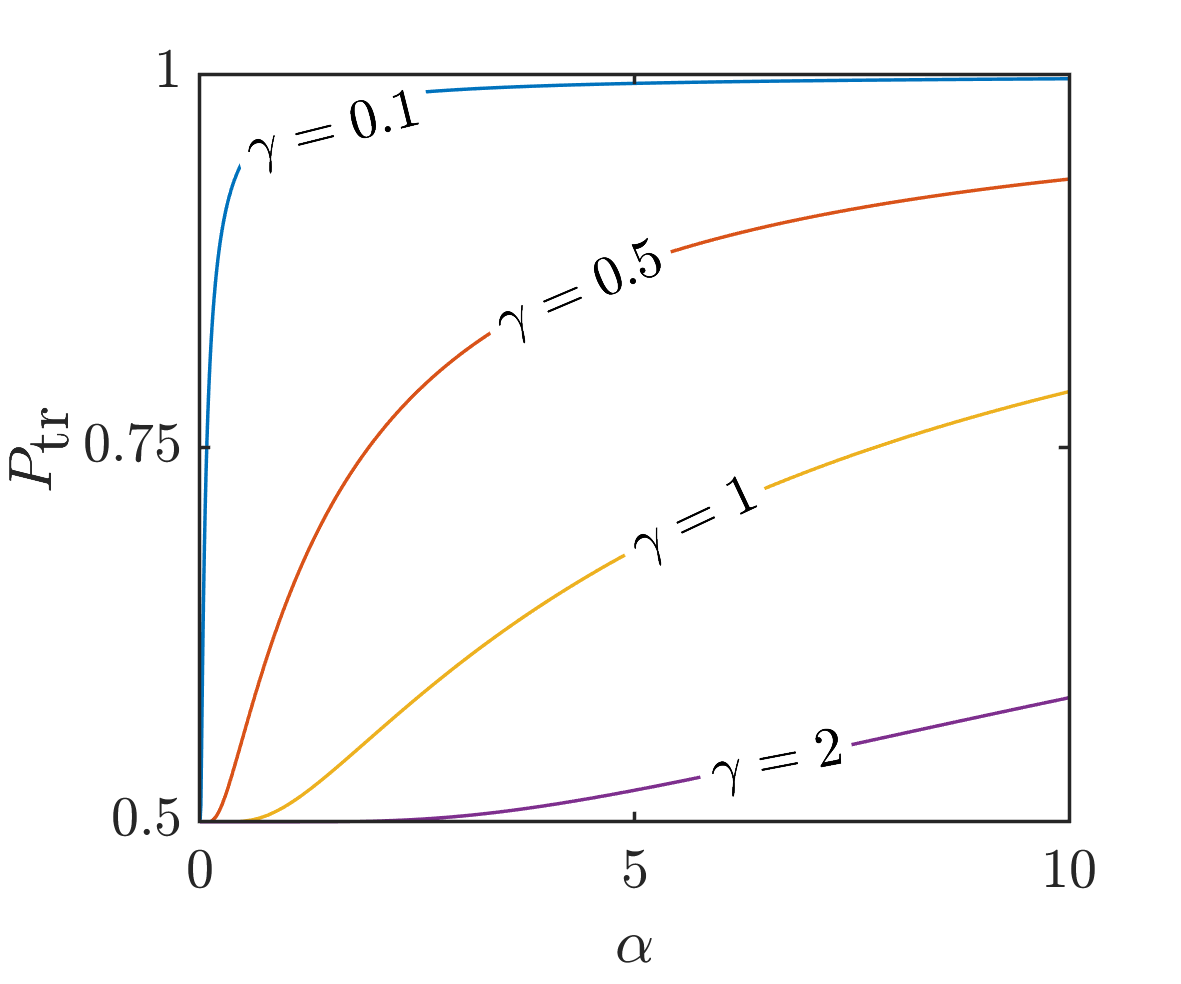}
\includegraphics[width=0.238\textwidth]{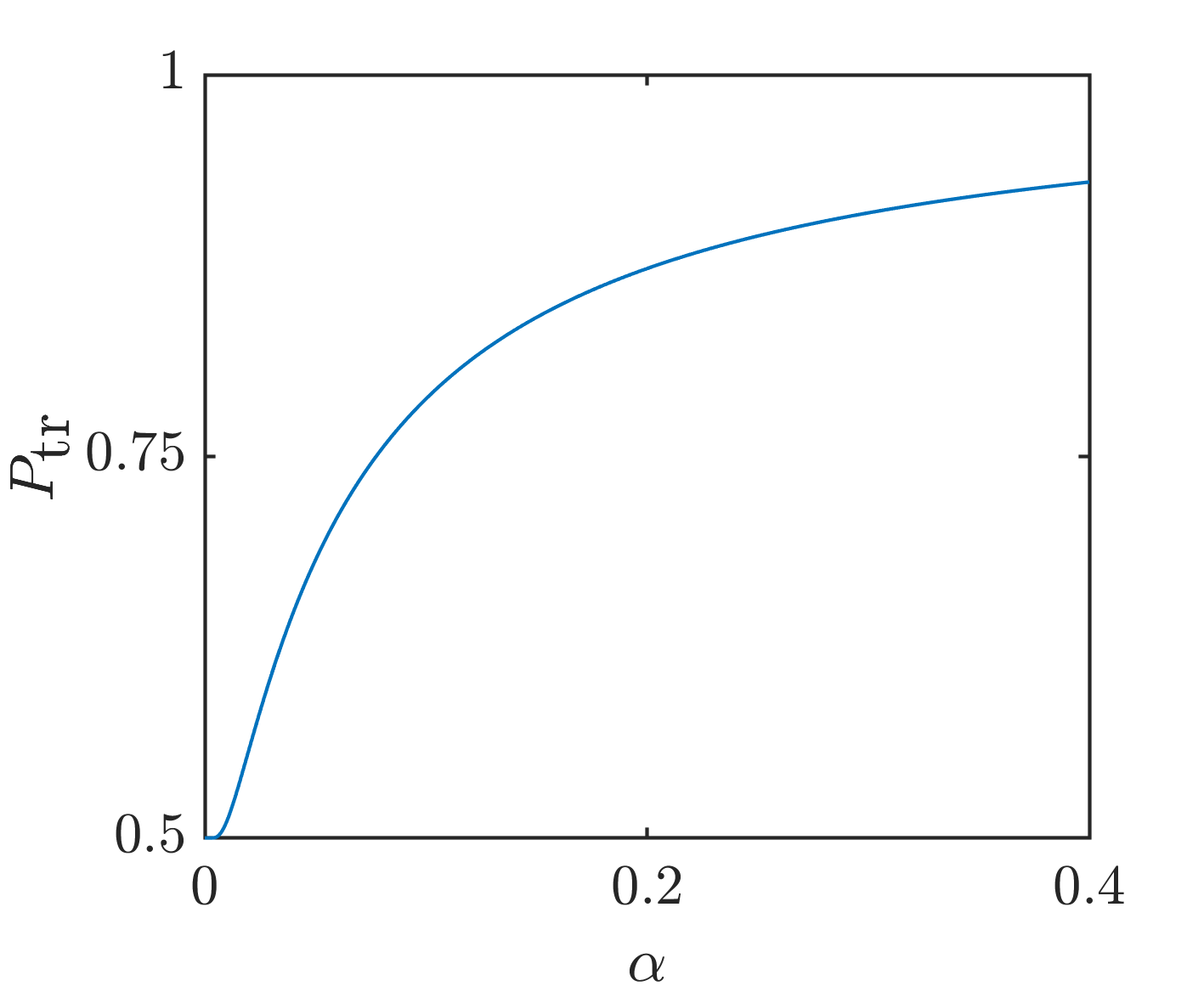}
\end{center}
\caption{\label{fig_LZ_prob} Transmission probability (\ref{eqn:lzprob}) as a function of the driving parameter. The left panel shows the variation with $\gamma$, the right panel focuses on small driving parameter values for $\gamma = 0.1$.}
\end{figure}

For unitary time evolution $|\psi_2\left(t \to +\infty\right)|^2$ can be obtained from $|\psi_1\left(t \to +\infty\right)|^2$ due to the conservation of probability. However, for the non-Hermitian dynamics considered here the total probability is no longer conserved. The amplitude $|\psi_2\left(t \to +\infty\right)|^2$ must be calculated by other means, starting with equation (\ref{eqn:psi2 eom}) and following a similar procedure to the one just used. 
Performing the transformation $z(t) = \ue^{i\pi/4} \sqrt{2\alpha}t$ converts the equation of motion for the component $\psi_2$
\begin{equation}
\label{eqn:psi2 second}
\ddot{\psi}_2 + \left(-\gamma^2 + i\alpha + \left(\alpha t\right)^2\right)\psi_2 = 0
\end{equation}
into the Weber equation
\begin{equation}
\label{eqn:psi2 weber}
\frac{d^2 \psi_2}{dz^2} + \left(-\nu + \frac{1}{2} - \frac{1}{4}z^2\right)\psi_2 = 0,
\end{equation}
where $\nu = -i\beta$ with $\beta = \gamma^2/2\alpha$. Due to the initial condition $|\psi_2\left(t \to -\infty\right)|^2 = 1$ the solution should be non-vanishing as $t \to -\infty$. Furthermore, the solution should have asymptotic behaviour consistent with equation (\ref{eqn:phi2 asymp m}). Thus, the solution is of the form $\psi_2(t) = B D_{-\nu}\left(-z(t)\right)$. Making use of the asymptotic expansion 
\begin{equation}
\label{eqn:phi2 asymp m2}
\psi_2\left(t \to -\infty\right) = B \ue^{-i\pi \nu/4}\ue^{-iR^2/4}R^{-\nu}
\end{equation}
and the initial conditions (\ref{eqn:boundary conds}) yields the normalisation factor $|B|^2 = \ue^{\pi \beta /2}$. This, together with the asymptotic result for the Weber function
\begin{equation}
\label{eqn:asymp weber}
\left|D_{-\nu}\left(-z(t\to+\infty)\right)\right|^2 = \ue^{3\pi\beta/2},
\end{equation}
leads to the amplitude
\begin{equation}
\label{eqn:phi2 mod asymp}
|\psi_2\left(t \to +\infty\right)|^2 = \ue^{2\pi \beta}.
\end{equation}
Inserting (\ref{eqn:phi1 mod asymp}) and (\ref{eqn:phi2 mod asymp}) into the definition of the transmission probability (\ref{eqn:landau zener def}) finally yields
\begin{equation}
\label{eqn:lzprob}
P_{\rm tr} = \left(2 - \ue^{-\frac{\pi \gamma^2}{\alpha}}\right)^{-1}.
\end{equation}
As expected, this approaches $\frac{1}{2}$ in the adiabatic limit. On the other hand, in the limit of fast driving, the usual quantum quench behaviour is observed, i.e. $|\psi_1|^2=0$ and $|\psi_2|^2=1$ for all times. For intermediate values of $\alpha$ the transmission probability monotonically increases with $\alpha$, interpolating between the two limits. Figure \ref{fig_LZ_prob} depicts the transmission probability as a function of $\alpha$ for various values of $\gamma$. 

\section{The two-mode model as the Bloch Hamiltonian of a PT-symmetric tight binding lattice}
We now demonstrate that the model (\ref{eqn:LZHam}) accurately describes the band transitions in a PT-symmetric lattice with an applied static force. This allows for a direct observation of the transition through a series of exceptional points in dependence on the adiabatic parameter, which can be tuned via the static force. Let us consider a PT-symmetric tight-binding Hamiltonian of the form
\begin{equation}
\begin{split}
\label{eqn_lattice_Ham}
\hat H =& \sum_{j}-\big(|j+1\rangle \langle j| \!+\! |j\rangle \langle j+1|\big) +i\Gamma (-1)^j |j \rangle\langle j|,
\end{split}
\end{equation}
with a gain and loss rate $\Gamma \in \R^+$. This model can be realised, for example, as a chain of waveguides with absorption in every other waveguide and optical gain of an equal strength in the waveguides in between. Some of the properties of this model have previously been discussed in \cite{Long09,Turk16}. A passively PT-symmetric version of this model has been implemented experimentally in \cite{Xu16b}. 

\begin{figure}
\begin{center}
\includegraphics[width=0.23\textwidth]{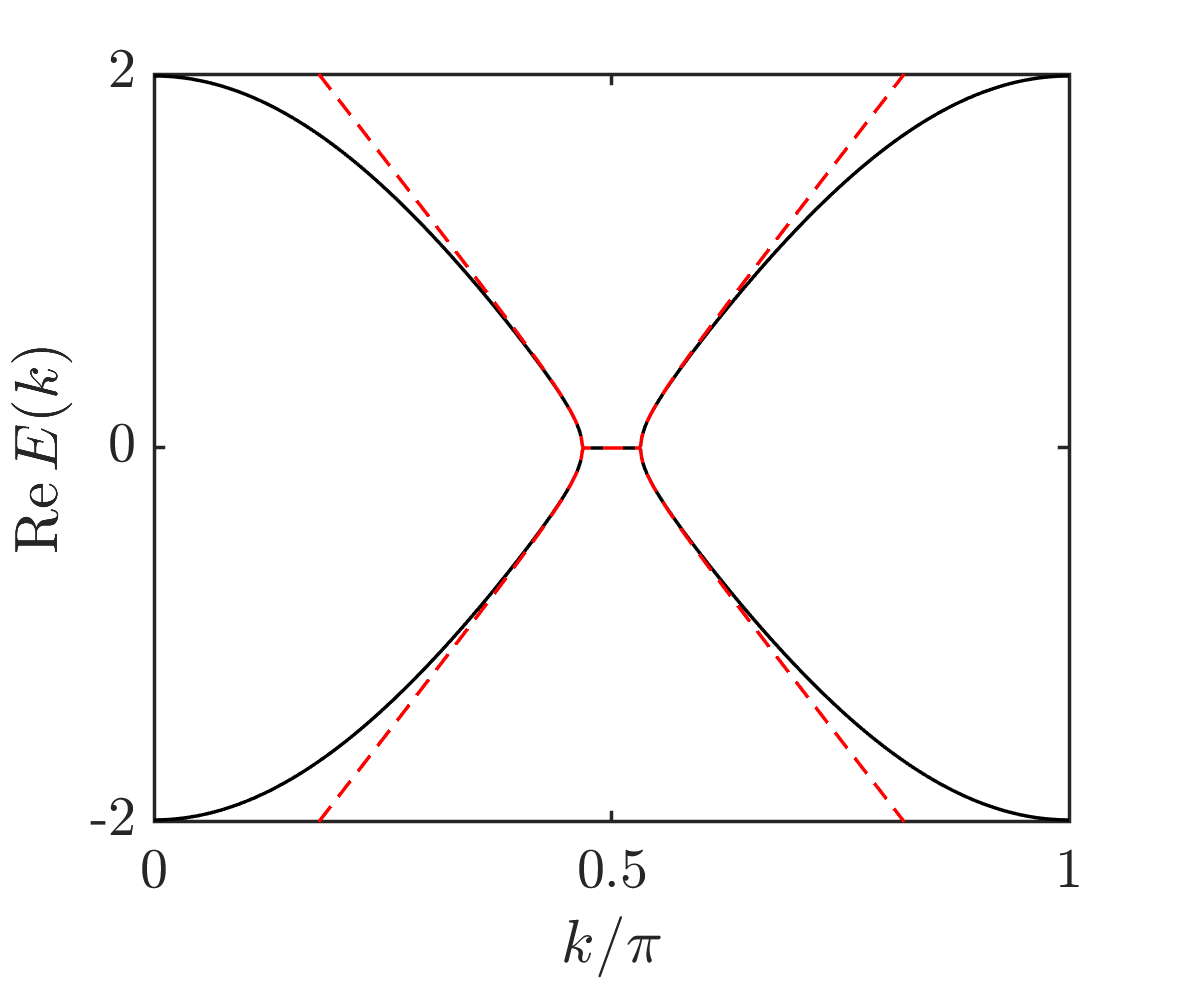}
\includegraphics[width=0.23\textwidth]{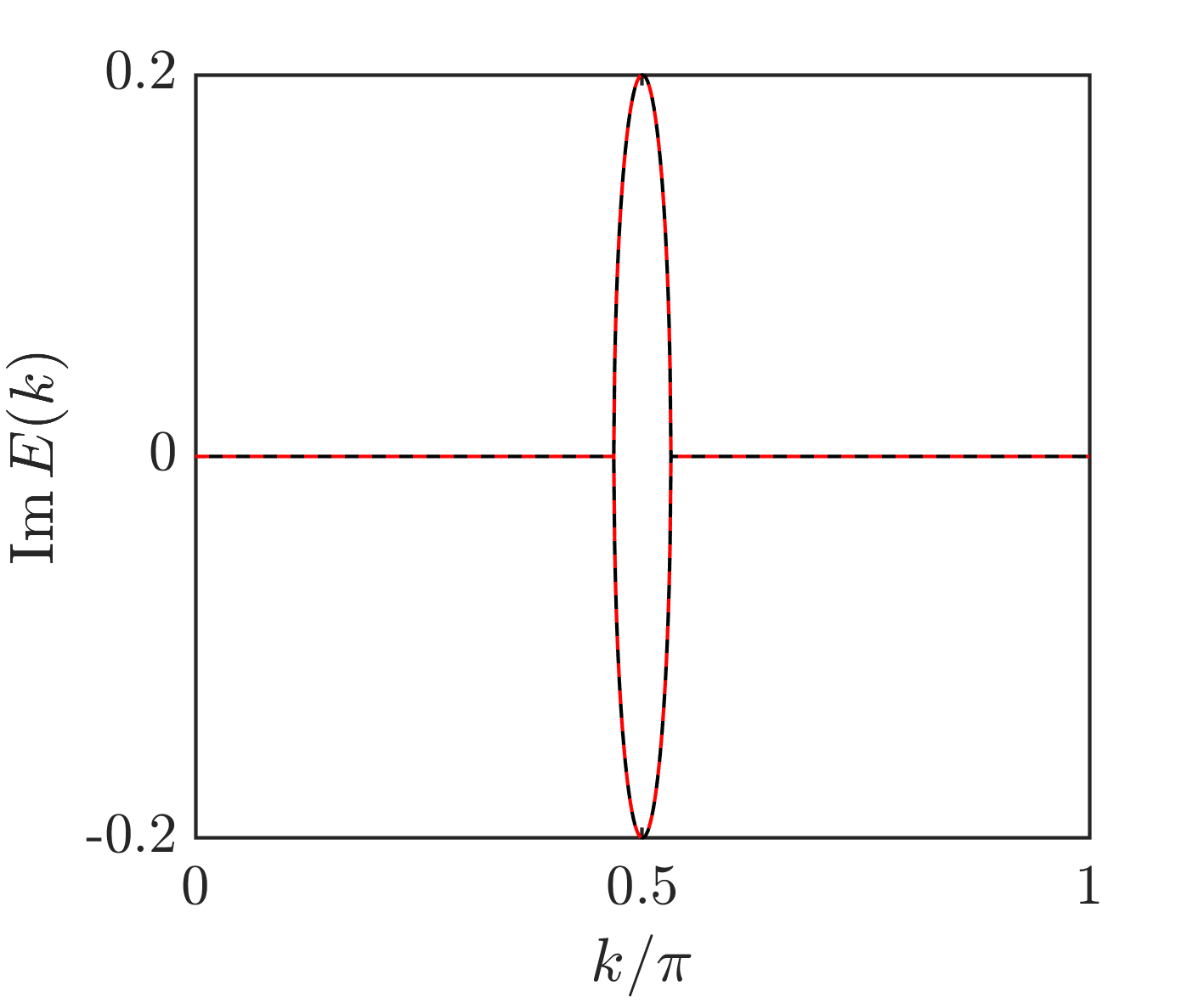}
\includegraphics[width=0.23\textwidth]{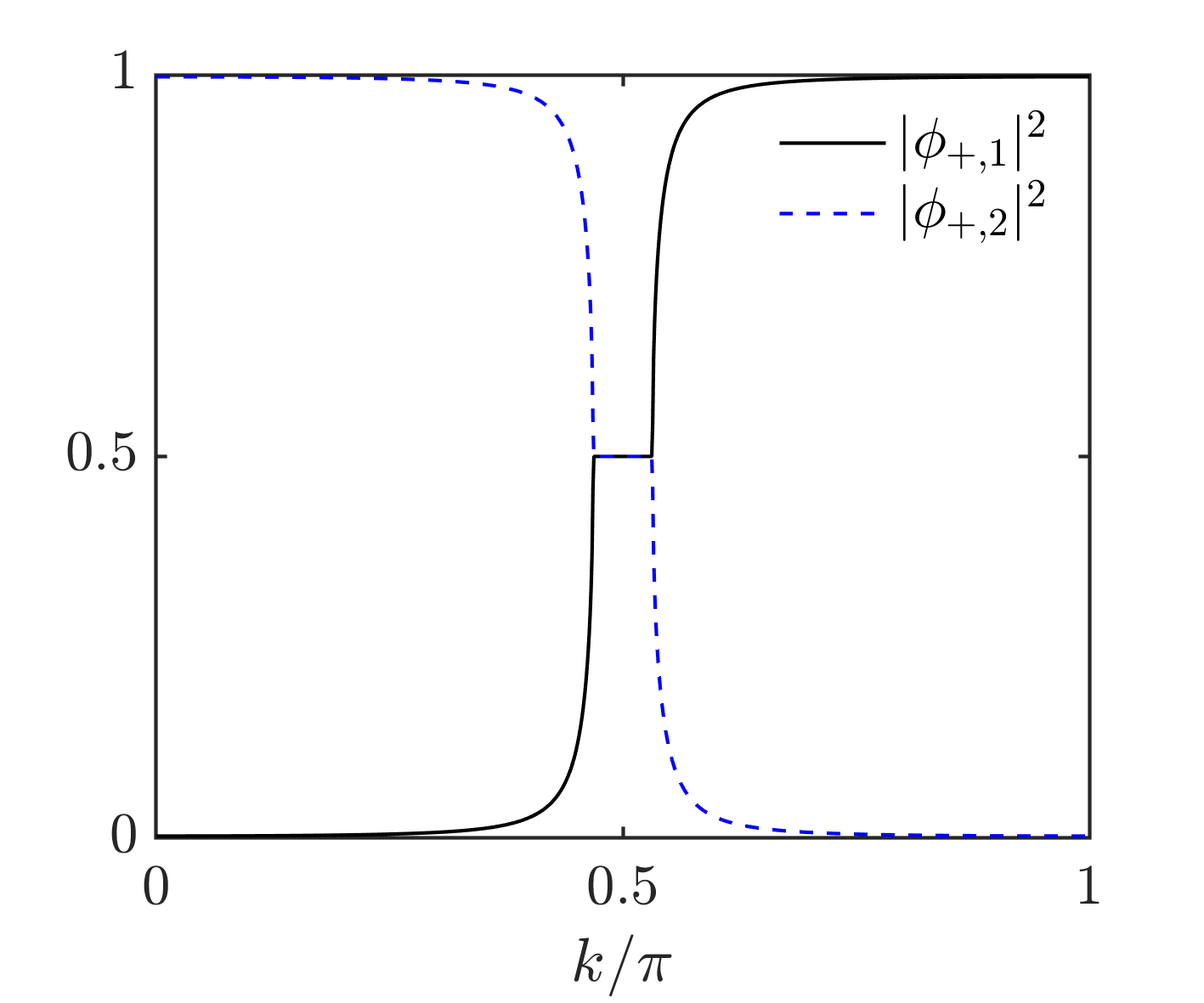}
\includegraphics[width=0.23\textwidth]{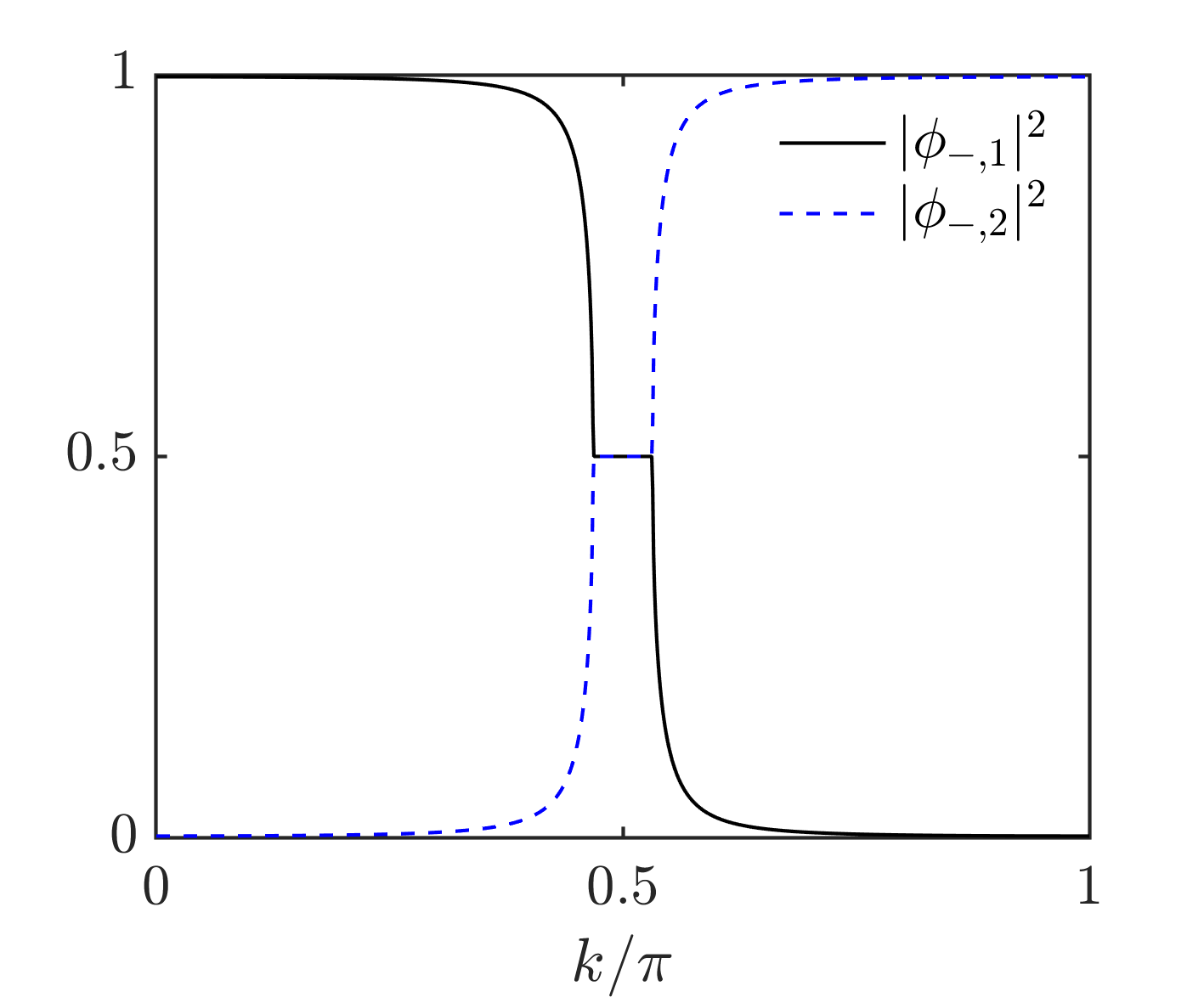}
\end{center}
\caption{\label{fig_eigvalBloch} Top row: Real (left) and imaginary (right) parts of the dispersion relation (\ref{eqn:dispersion}) for $\Gamma=0.2$. The red dashed line depicts the same quantities for the Taylor expansion of the Bloch Hamiltonian (\ref{eqn:field free ham}) around $k = \pi/2$. Bottom row: Modulus squared of the components of the eigenstates $\phi_{\pm}$ associated with $E_{\pm}(k)$.}
\end{figure}

\begin{figure}
\begin{center}
\includegraphics[width=0.23\textwidth]{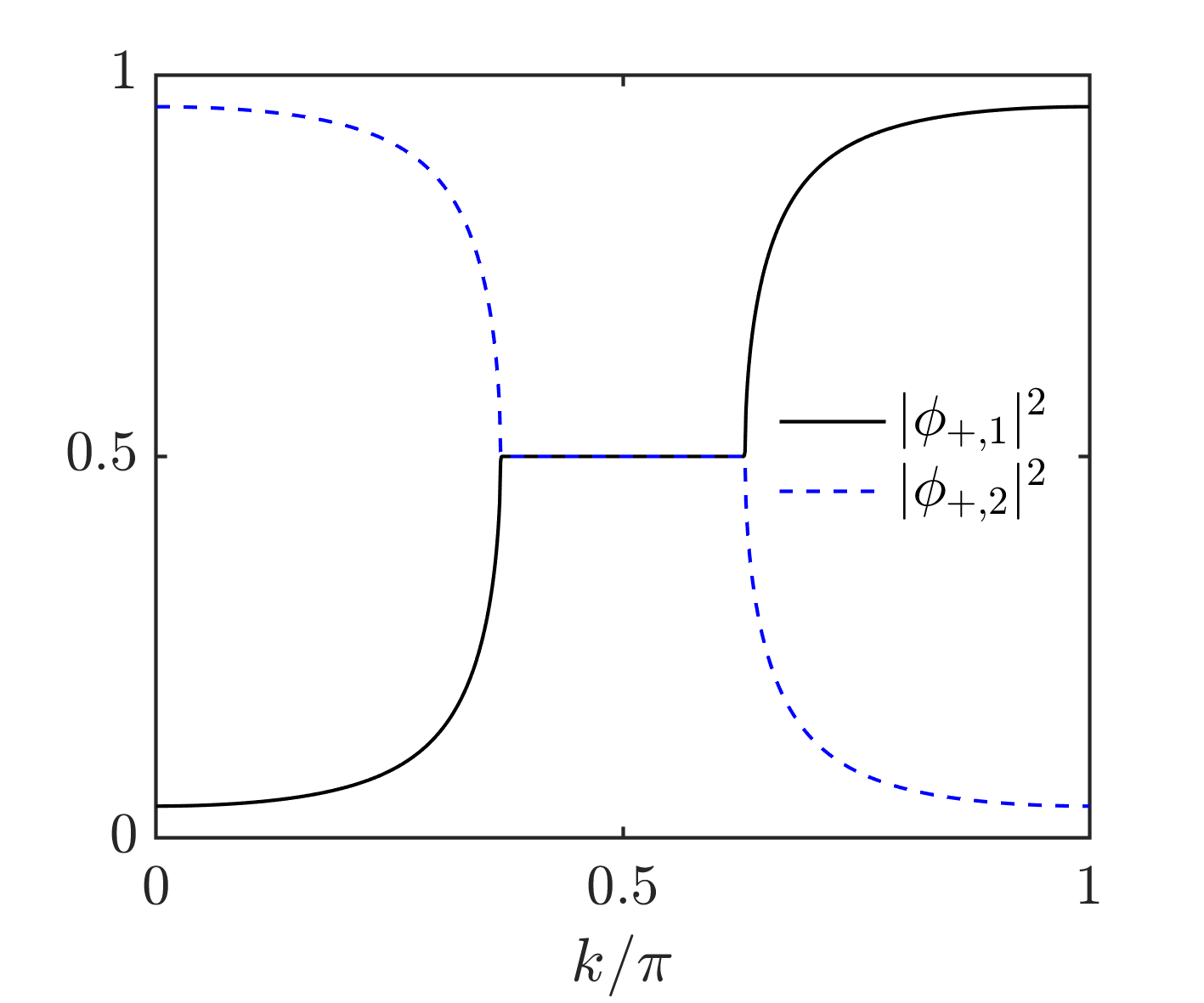}
\includegraphics[width=0.23\textwidth]{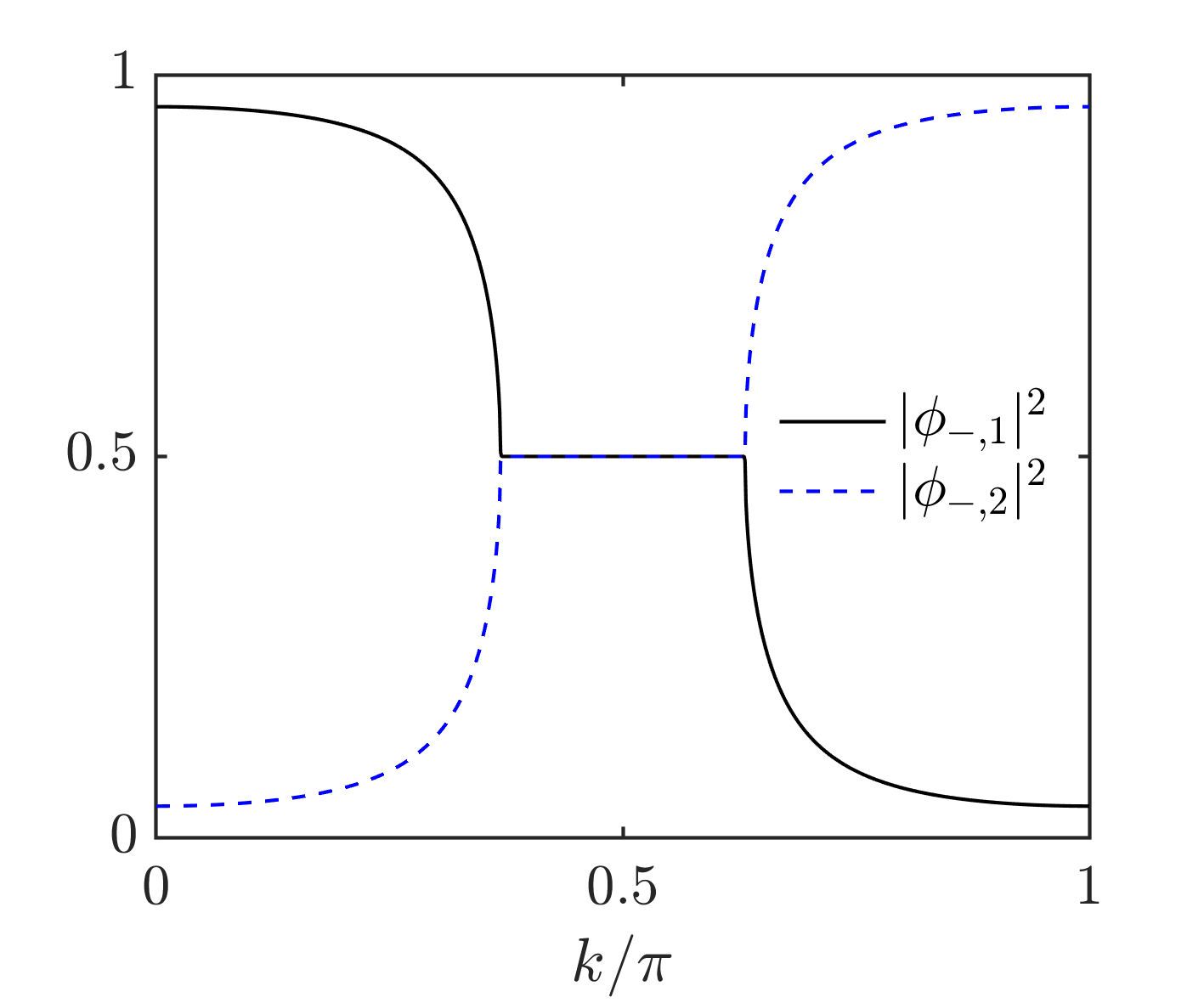}
\end{center}
\caption{\label{fig_eigvalBloch2} Modulus squared of the components of the eigenstates $\phi_{\pm}$ associated with $E_{\pm}(k)$ for $\Gamma = 0.8$.}
\end{figure}

It is convenient to study the system in the quasimomentum representation, that is, in the basis of the Bloch states
\begin{equation}
\label{eqn:bloch states}
|k\ra = \sum_j |j\ra\la j|k\ra = \frac{1}{\sqrt{2\pi}} \sum_j \ue^{ikj}|j\ra,
\end{equation}
where the quasimomentum $k$ is confined to the region $-\pi \leq k \leq \pi$. The Bloch states are orthogonal and normalised to the $2\pi$-periodic delta comb $\la k|k'\ra = \delta_{2\pi}\left(k'-k\right)$. In a similar spirit to \cite{Long10} we introduce the two-component function $\Psi = \left(\Psi_1,\Psi_2\right)$, with $\Psi_1(k) = \psi(k)$ and $\Psi_2(k) = \psi(k+\pi)$. The time evolution may then be written as the two-level Schr\"odinger equation
\begin{equation}
\label{eqn:spinor eom}
i\dot{\Psi} = \left(-2\cos k \sigma_z +  i\Gamma \sigma_x\right)\Psi =h(k) \Psi,
\end{equation}
where $\sigma_i$ are Pauli matrices and the Bloch Hamiltonian is defined as
\begin{equation}
\label{eqn:field free ham}
h(k) = \begin{pmatrix} -2\cos k  &i\Gamma\\ i\Gamma & 2\cos k \end{pmatrix}.
\end{equation}
The $k$-dependent eigenvalues of $h(k)$ define the dispersion relation of the two band system
\begin{equation}
\label{eqn:dispersion}
E_\pm(k) =  \pm \sqrt{4 \cos^2 k - \Gamma^2}.
\end{equation}
The band structure is complex for arbitrarily small values of $\Gamma$. For values of $\Gamma< 2$ there are exceptional points at $2 |\cos k| = \Gamma$ and when $2|\cos k| < \Gamma$ the energy is imaginary. For $\Gamma>2$ the bands are purely imaginary and there are no exceptional points. Here we focus on $\Gamma$ values well below this critical point. An example of the band structure for $\Gamma=0.2$ is depicted in the top row of Fig. \ref{fig_eigvalBloch}. The bottom row of the same figure shows the modulus squared of the components of the eigenstates. The eigenstates are close to the standard basis vectors at $k=0$ and $k=\pi$. However, when $\Gamma$ is increased this is no longer the case, as illustrated in Fig. \ref{fig_eigvalBloch2} for $\Gamma = 0.8$.

We now show that the two-component function $\Psi(k)$ of a broad Gaussian beam in position space is an approximate eigenstate of the Bloch Hamiltonian, when the Bloch states are close to the standard basis vectors. Consider a Gaussian in real space representation given by
\begin{equation}
\psi(j) = \mathcal{N}\ue^{-(j-q_0)^2/2\sigma^2 + ik_0(j-q_0)},
\end{equation}
where $j$ is the lattice index, $q_0$ is the centre of the Gaussian, $k_0$ is the initial momentum, $\sigma$ is the width parameter and $\mathcal{N}$ is a normalisation constant, chosen so that $\sum_j |\psi(j)|^2 = 1$.
The quasimomentum representation of this state is found to be
\begin{equation}
\label{eqn:psi k jacob}
\psi(k) = \frac{\sqrt{\sigma}}{\pi^{1/4}} \frac{\theta_3\left(z,\ue^{i\pi\tau}\right)}{\sqrt{\theta_3\left(-q_0 \pi,\ue^{i\pi\tau/2}\right)}}\,\ue^{-\frac{\sigma^2}{2}\left(k-k_0\right)^2 - iq_0 k},
\end{equation}
where $z = i\sigma^2\pi(k-k_0)-q_0 \pi$, $\tau = 2i\sigma^2\pi$ and the Jacobi theta function $\theta_3(z,\ue^{i\pi\tau})$ is defined as \cite{Abra72}
\begin{equation}
\label{eqn:jacobi 3}
\theta_3\left(z,\ue^{i\pi\tau}\right) = 1 + 2\sum_{n=1}^\infty \ue^{i\pi\tau n^2}\cos \left(2nz\right).
\end{equation}
Thus, $\psi(k)$ is the product of a Gaussian distribution in quasimomentum space and a term involving Jacobi theta functions. Using the properties of the theta function it is straightforward to check that $\psi(k+2m\pi) = \psi(k)$ for any integer $m$. 

In order to obtain the two-component function $\Psi(k)$ we need to calculate $\psi(k+\pi)$. From (\ref{eqn:psi k jacob}) it follows that 
\begin{equation}
\label{eqn:psi k jacob pi}
\psi(k+\pi) = \frac{\sqrt{\sigma}}{\pi^{1/4}} \frac{\theta_2\left(z,\ue^{i\pi\tau}\right)}{\sqrt{\theta_3\left(-q_0 \pi,\ue^{i\pi\tau/2}\right)}}\, \ue^{-\frac{\sigma^2}{2}\left(k-k_0\right)^2 - iq_0 k},
\end{equation}
where the theta function $\theta_2(z,\ue^{i\pi\tau})$ is defined as
\begin{equation}
\label{eqn:jacobi 2}
\theta_2(z,\ue^{i\pi\tau}) = 2 \ue^{i\pi\tau/4}\sum_{n=0}^\infty \ue^{i\pi\tau n(n+1)}\cos\left((2n+1)z\right)
\end{equation}
and we made use of the relationship
\begin{equation}
\theta_2(z,\ue^{i\pi\tau}) = \ue^{iz + i\pi\tau/4}\theta_3\left(z+ \frac{1}{2}\pi\tau,\ue^{i\pi\tau}\right).
\end{equation}
Thus, $\Psi(k)$ may be written as
\begin{equation}
\label{eqn:init 2 comp}
\Psi(k) = \frac{\sqrt{\sigma}}{\pi^{1/4}} \frac{\ue^{-\frac{\sigma^2}{2}\left(k-k_0\right)^2 - iq_0 k}}{\sqrt{\theta_3\left(-q_0 \pi,\ue^{i\pi\tau/2}\right)}}\begin{pmatrix} \theta_3\left(z,\ue^{i\pi\tau}\right) \\ \theta_2\left(z,\ue^{i\pi\tau}\right) \end{pmatrix}.
\end{equation}
In the broad Gaussian limit $\sigma \to \infty$, $\tau$ tends to infinity along the imaginary axis and the two-components become $(\theta_3(z,\ue^{i\pi\tau}),\theta_2(z,\ue^{i\pi\tau})) \to (1,0)$. The two-component function $\Psi(k)$ then becomes a standard basis vector multiplied by a Gaussian wave packet that is highly localised around $k_0$. So, for example, a broad Gaussian beam in position space with momentum $k_0=0$ ($k_0=\pi$) yields a $\Psi(k)$ that is approximately one of the eigenstates of the Bloch Hamiltonian at $k=0$ ($k=\pi$) depicted in Fig. \ref{fig_eigvalBloch}. 

\begin{figure}
\begin{center}
\includegraphics[width=0.23\textwidth]{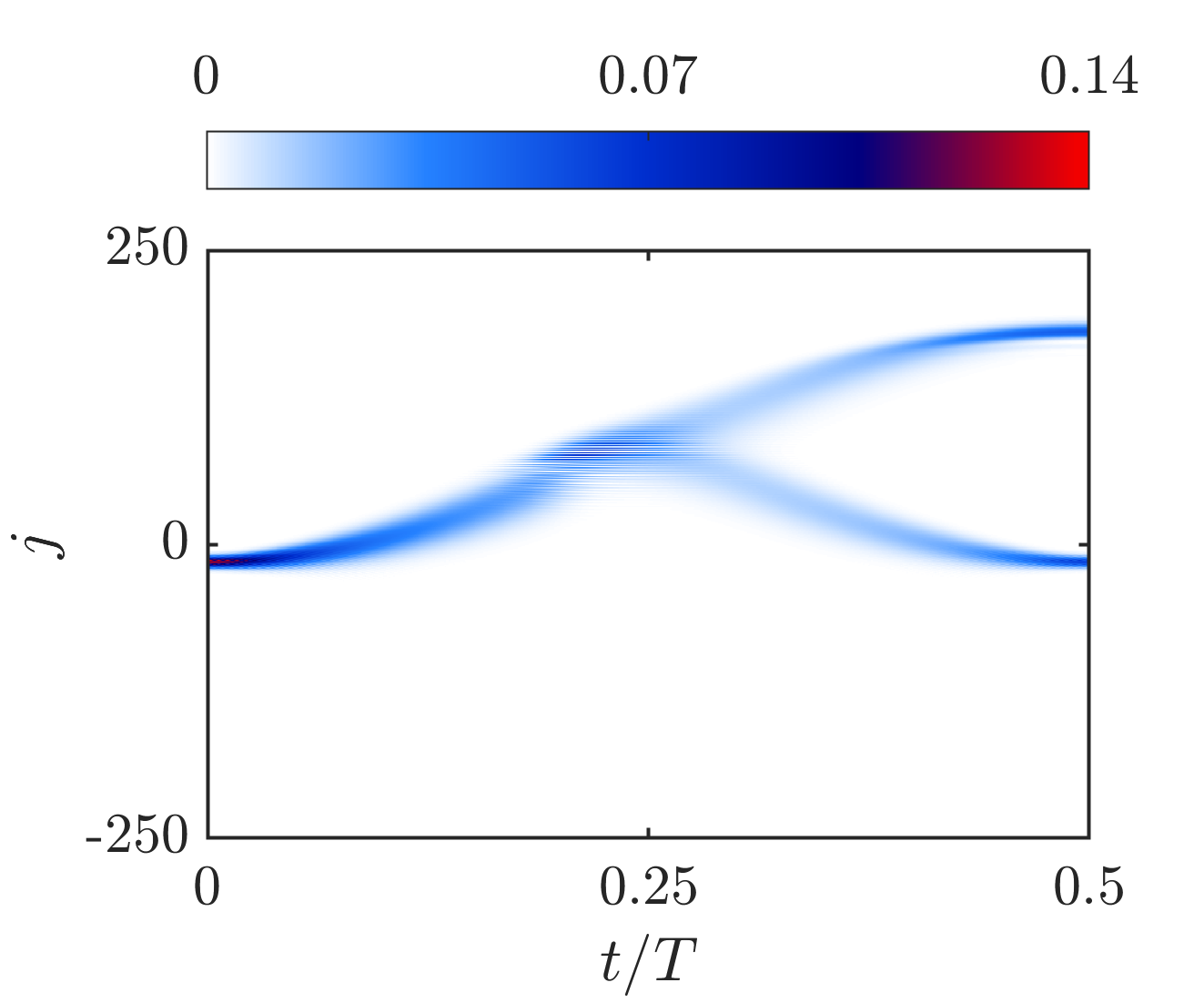}
\includegraphics[width=0.23\textwidth]{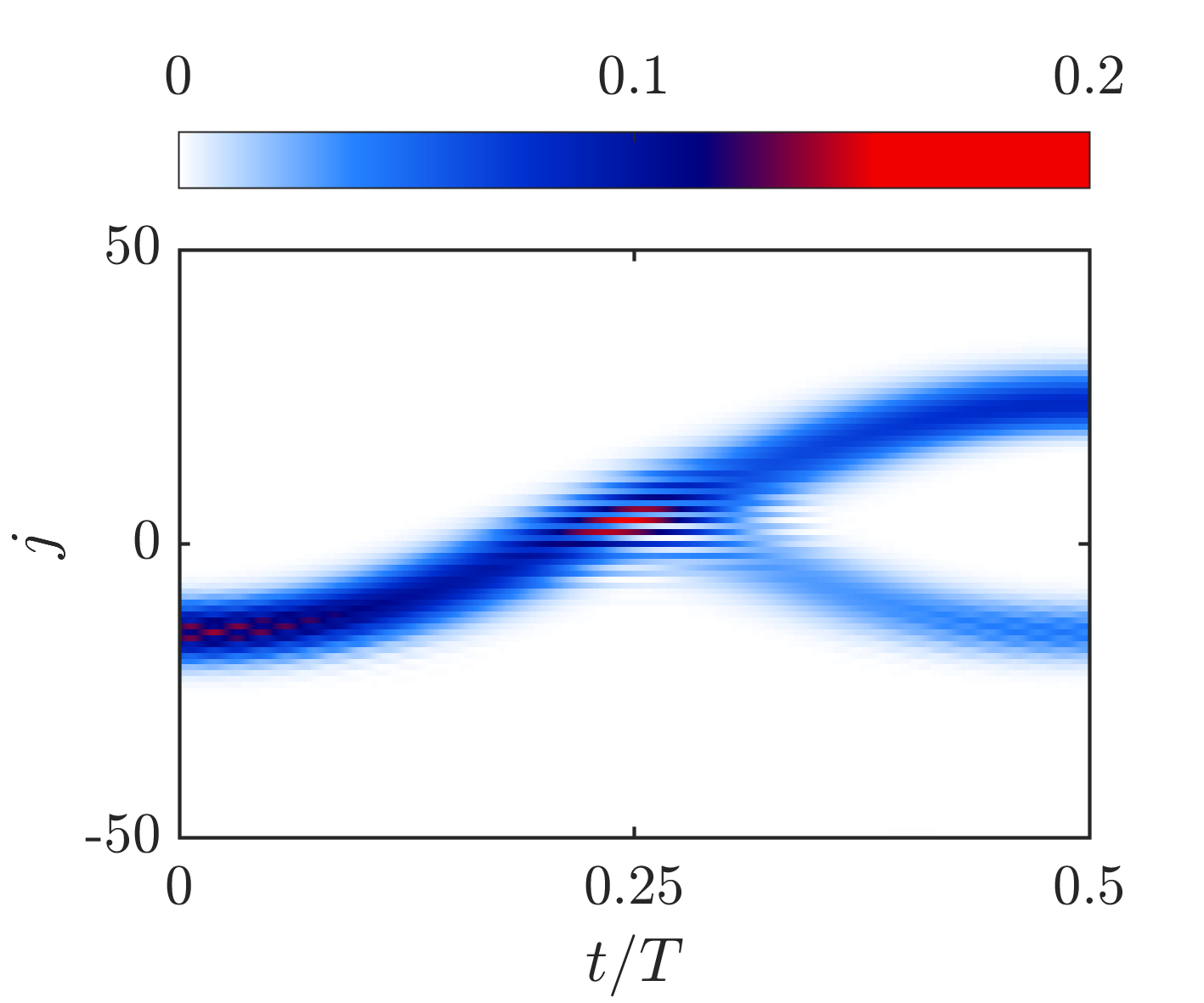}
\includegraphics[width=0.23\textwidth]{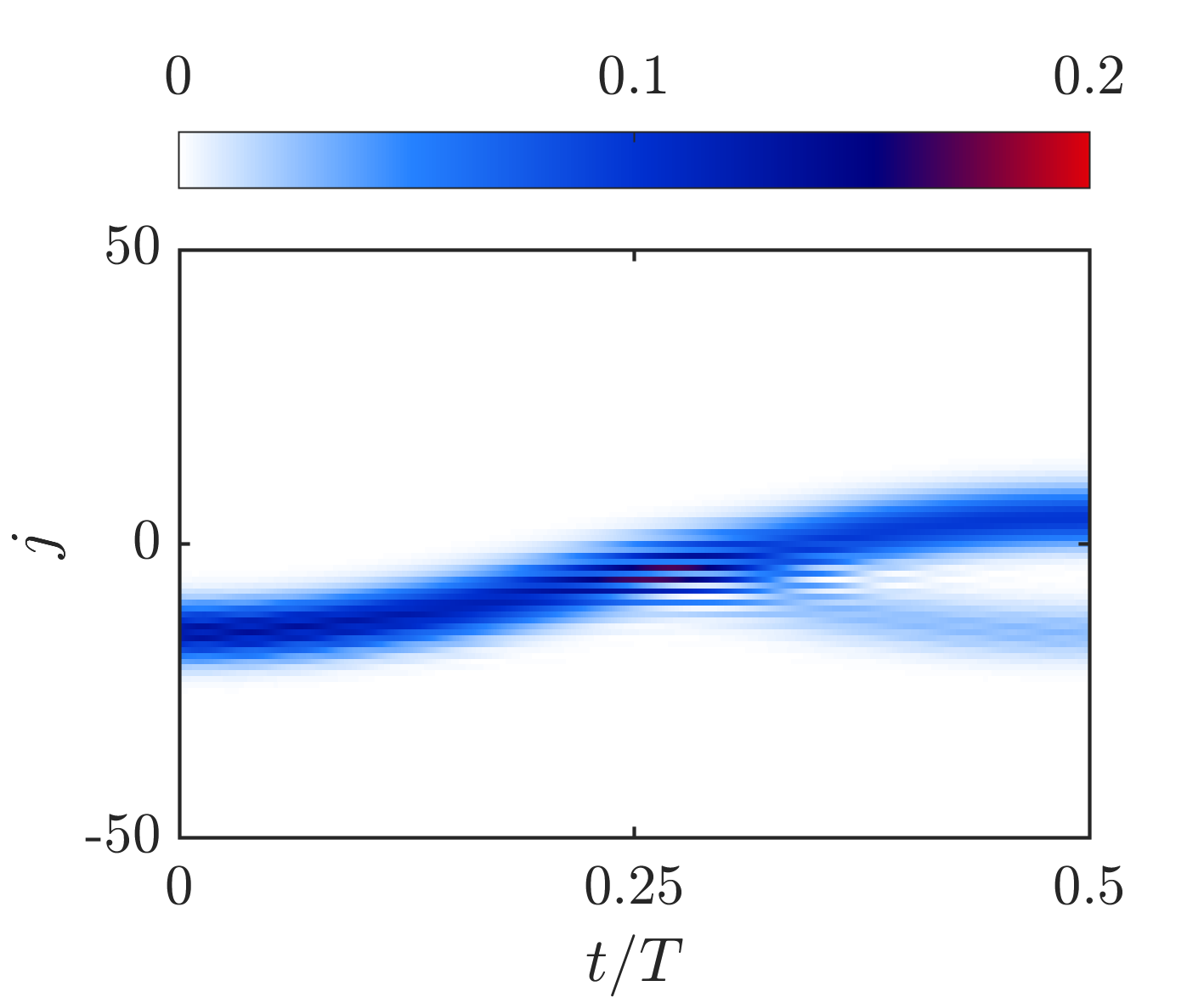}
\includegraphics[width=0.23\textwidth]{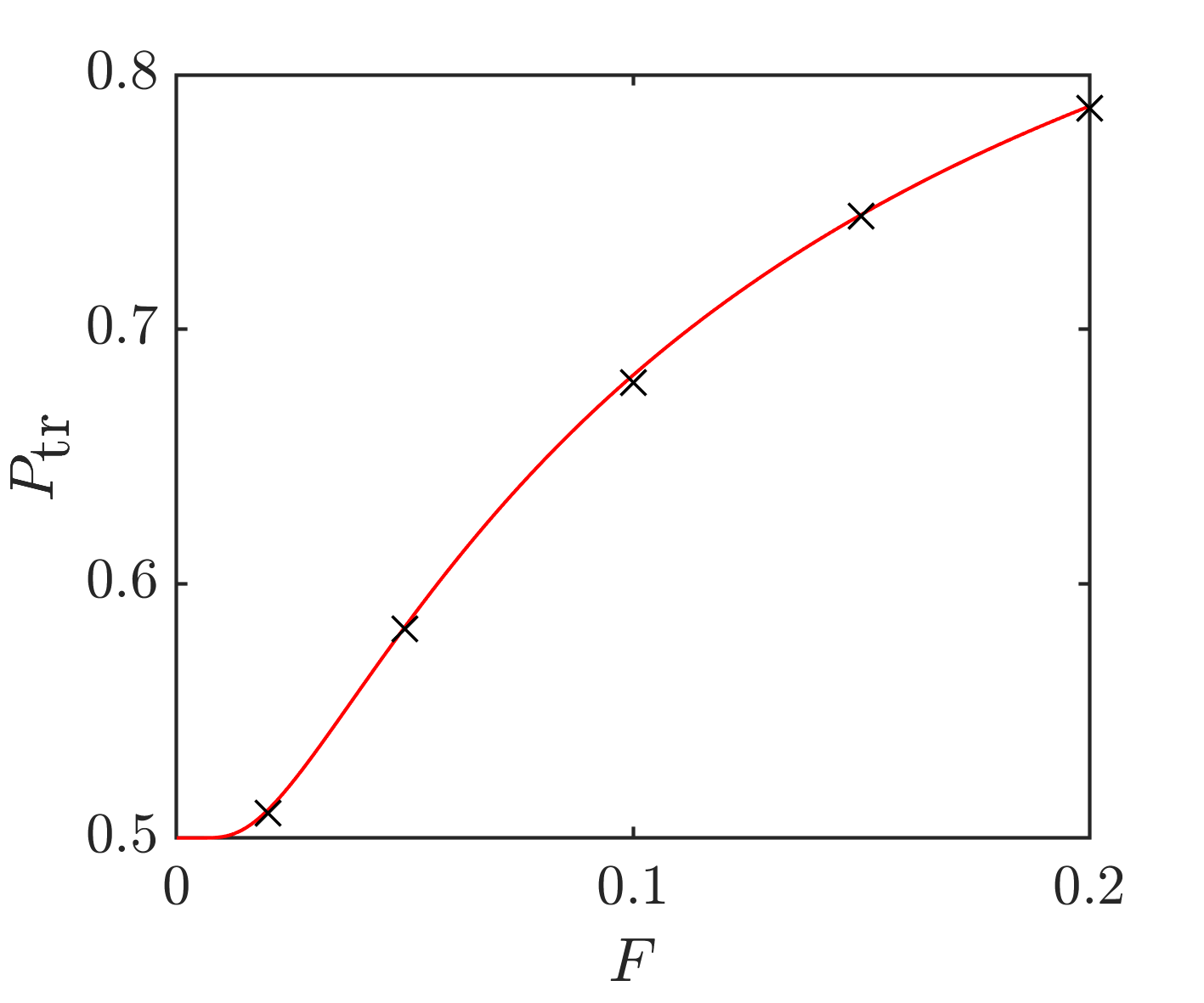}
\end{center}
\caption{\label{fig_lz_comparison} Time evolution of the renormalised density for an initial Gaussian wave packet in position space with centre $x_0 = -15$, momentum $k_0 = \pi$ and width $\sigma^2 = 20$, for $\Gamma = 0.2$ and different values of $F$ ($F=0.02$ top left, $F=0.1$ top right, and $F=0.2$ bottom left). The bottom right panel shows the transmission probability for $\Gamma = 0.2$ as a function of $F$ (red). The crosses are numerical values of the transmitted fraction, obtained by evaluating the population in the upper branch of the beam in position space at $t = T/2$, where the Bloch period $T=2\pi/F$.}
\end{figure}

If a static force is applied to the lattice a term $F \sum_{j} j |j\rangle\langle j|$ is added to the Hamiltonian (\ref{eqn_lattice_Ham}). An initial state $\Psi(k)$ that is approximately an eigenstate of the Bloch Hamiltonian will then perform a non-Hermitian version of the famous Bloch oscillations. This is accompanied by transitions between the bands, which appear as a splitting of the beam in real space. Some examples of the resulting dynamics can be seen in Fig. \ref{fig_lz_comparison}. Similar behaviour has been observed experimentally in a system of optical fibre loops \cite{Wimm15}. The populations of the two bands, i.e. the relative amplitudes of the two beams, are approximated by the Landau-Zener-type Hamiltonian (\ref{eqn:LZHam}). This can be understood in the following way. 

The static force introduces a term $Fq$ into the two-component Hamiltonian (\ref{eqn:spinor eom})
\begin{equation}
\label{eqn:full spin ham}
h(k,q) = -2\cos k \sigma_z +  i\Gamma \sigma_x + Fq,
\end{equation}
where $q = id/dk$ is canonically conjugate to $k$ with $[q,k] = i$. The expectation value of $k$ evaluated in the two-component state $\Psi(k,t)$ is
\begin{equation}
\la k \ra_t = \frac{\int dk\, \Psi^\dag(k,0) U^\dagger k U \Psi(k,0)}{\int dk\, \Psi^\dag(k,0) U^\dagger U \Psi(k,0)},
\end{equation}
where we have defined the (non-unitary) time-evolution operator $U=\ue^{-i h(k,q) t}$ and the integrals are over the interval $\left[-\pi/2,\pi/2\right]$. The Zassenhaus formula enables the factorisation $U=\ue^{A(k)}\ue^{-i F q t}$, with some matrix operator $A$ that is independent of $q$, such that
\begin{equation}
\label{eqn:k expec spinor}
\la k \ra_t = \frac{\int dk\, k\Psi^\dag(k+Ft,0)\ue^{A^\dagger(k)}\ue^{A(k)} \Psi(k+Ft,0)}{\int dk\, \Psi^\dag(k+Ft,0)\ue^{A^\dagger(k)}\ue^{A(k)} \Psi(k+Ft,0)}.
\end{equation}
For a Gaussian in position space the two-component function $\Psi(k,0)$ is given by equation (\ref{eqn:init 2 comp}), and in the broad beam limit the expectation value (\ref{eqn:k expec spinor}) reduces to
\begin{equation}
\la k \ra_t = \langle k\rangle_0-Ft.
\end{equation}
This is the acceleration theorem that is well known for Hermitian systems. In the non-Hermitian case this is only an approximation and relies on the initial quasimomentum uncertainty being negligible. It follows from the properties of the delta function that $\la k \ra_0 = 0$ when $k_0=0$ or $k_0 = \pi$. 

In summary, if a static force is applied to the PT-symmetric chain (\ref{eqn_lattice_Ham}), and $\Psi(k,0)$ is approximately a Bloch state, then the dynamics can be described by the effectively time-dependent Bloch Hamiltonian (\ref{eqn:field free ham}), with the operator $k$ replaced by its time-dependent expectation value $\la k \ra_t = \langle k\rangle_0 - Ft$. We can further 
Taylor expand the effective two-level system (\ref{eqn:field free ham}) around the band edge $k = \pi/2$. The eigenvalues of the Taylor expanded Hamiltonian are depicted in the top panel of Fig. \ref{fig_eigvalBloch}, in comparison to the exact eigenvalues for $\Gamma=0.2$. We observe a good agreement of the eigenvalues close to the exceptional points. The resulting Hamiltonian is of the form (\ref{eqn:LZHam}), with $\alpha \to 2/F$ and $\gamma \to \Gamma/F$, and the instantaneous eigenstates represent the two quasimomentum bands of the system. It follows from (\ref{eqn:lzprob}) that the transmission probability is
\begin{equation}
\label{eqn:lzprob_model}
P_{\rm tr} = \left(2 - \ue^{-\frac{\pi\Gamma^2}{2F}}\right)^{-1}.
\end{equation}
Thus, if $\Psi(k,0)$ is initialised in an approximate eigenstate of the Bloch Hamiltonian at $k=0$ or $k=\pi$, then after half a Bloch period $t = T/2$ (with $T=2\pi/F$) the populations of the two bands are approximated by (\ref{eqn:lzprob_model}).

This result agrees well with numerical calculations, as demonstrated in Fig. \ref{fig_lz_comparison} for an initial broad Gaussian beam. The time evolution of the renormalised density $|\psi_j|^2/|\psi|^2$ on each site $j$ is plotted for $\Gamma=0.2$ and three values of $F$. We observe that the Bloch oscillations sweep through the band structure depicted in Fig. \ref{fig_eigvalBloch}. The data points in the bottom right panel were obtained by evolving the same initial state up to half the Bloch period for various values of $F$. Each point corresponds to the population in the upper branch of the beam in position space at $t=T/2$ for a particular $F$ value. The agreement between the numerically observed transmission probability and the approximative formula (\ref{eqn:lzprob_model}) is excellent. This behaviour is observed for a wide range of parameters, as long as the two beams splitting at the exceptional point can be meaningfully distinguished. We expect the experimental observation of this transition due to sweeping through exceptional points to be entirely within reach. 
 
\section{Summary}
We have investigated a two-level system driven through two consecutive exceptional points adiabatically and at finite speed. In the adiabatic limit this leads to behaviour having no analogue in the Hermitian case. The population is equally distributed between the states coalescing in the exceptional point, corresponding to a loss of information of the initial state. In the limit of fast driving the familiar quantum quench behaviour is recovered. We have derived an analytic expression for the population transfer for arbitrary speed of parameter variation, interpolating between these two extremes. We have further demonstrated how this can be experimentally investigated in a PT-symmetric lattice using Bloch oscillations, such as an optical waveguide setup, providing new opportunities for engineering beam dynamics. 

\section*{Acknowledgements}
The authors would like to express their thanks to Stefan Rotter and Alexander Schumer for a stimulating exchange regarding the loss of information in the adiabatic limit, and to Dorje Brody and Henning Schomerus for further useful discussions on this topic.

E.M.G.  acknowledges  support  from  the Royal Society (Grant. No. UF130339) and from the European Research Council (ERC) under the European Union's Horizon 2020 research and innovation programme (grant agreement No 758453). B.L. acknowledges support from the Engineering and Physical Sciences Research Council via the Doctoral Training Partnership (Grant No. EP/M507878/1).

\end{document}